\documentclass[12pt]{article}
\usepackage[english]{babel}
\usepackage{amsmath,amsthm}
\usepackage{amsfonts}
\usepackage{graphicx}
\usepackage{enumerate}
\usepackage[usenames,dvipsnames]{color}
\usepackage{subfigure}
\usepackage[right,pagewise,displaymath, mathlines]{lineno}
\usepackage{epstopdf}
\usepackage{color}

\numberwithin{equation}{section}
\numberwithin{figure}{section}
\numberwithin{table}{section}

\newtheorem{theorem}{\sc Theorem}[section]

\newtheorem{property}[theorem]{\sc Property}
\newtheorem{note}{\sc Note}[section]

\numberwithin{equation}{section}

\begin{document}

\begin{quote}

\begin{center}
{\bf \large
Measuring economic inequality and risk: a unifying approach based on personal gambles, societal preferences and references}
\vspace*{8mm}

\textsc{Francesca Greselin}$^{*}$
\smallskip

\textit{Dipartimento di Statistica e Metodi Quantitativi, Universit\`a di Milano--Bicocca, Milan, Italy}

\medskip

\textsc{and}

\medskip

\textsc{Ri\v{c}ardas Zitikis}
\smallskip

\textit{Department of Statistical and Actuarial Sciences,
University of Western Ontario,
London, Ontario N6A~5B7, Canada}

\medskip

$^{*}$Corresponding author: francesca.greselin@unimib.it

\end{center}

\medskip

\noindent
The underlying idea behind the construction of indices of economic inequality is based on measuring deviations of various portions of low incomes from certain references or benchmarks, that could be point measures like population mean or median, or curves like the hypotenuse of the right triangle where every Lorenz curve falls into. In this paper we argue that by appropriately choosing population-based references, called societal references, and distributions of personal positions, called gambles, which are random, we can meaningfully unify classical and contemporary indices of economic inequality, as well as various measures of risk. To illustrate the herein proposed approach, we put forward and explore a risk measure that takes into account the relativity of large risks with respect to small ones.

\medskip

\end{quote}

\newpage

\section{\normalsize Introduction}
\label{introduction}

The Gini mean difference and its normalized version, known as the Gini index, have aided decision makers since their introduction by Corrado Gini more than a hundred years ago (Gini, 1912, 1914, 1921; see also Giorgi, 1990, 1993; Ceriani \& Verme, 2012; and references therein). In particular, the Gini index has been widely used by economists and sociologists to measure economic inequality. Measures inspired by the index have been employed to assess the equality of opportunity (e.g., Weymark 2003; Kovacevic, 2010; Roemer, 2013) and estimate income mobility (e.g., Shorrocks, 1978). Policy makers have used the Gini index in quantitative development policy analysis (e.g., Sadoulet \& de Janvry, 1995) and in particular for assessing the impact of carbon tax on income distribution (e.g., Oladosu \& Rose, 2007). The index has been employed for analysing inequality in the use of natural resources (e.g., Thompson, 1976) and for developing informed policies for sustainable consumption and social justice (e.g., Druckman \& Jackson, 2008). Various extensions and generalizations of the idex have been used to evaluate social welfare programs (e.g., Duclos, 2000; Kenworthy \& Pontusson, 2005; Korpi \& Palme, 1998; Ostry et al., 2014) and to improve the knowledge of tax-base and tax-rate effects as well as of temporal repercussions of distinct patterns of taxation and public finance on the society (e.g., Pf\"{a}hler, 1990; Slemrod, 1992; Yitzhaki, 1994; Van De Ven et al., 2001). Several classification methods (e.g., Hand, 2001) have employed the Gini index for modeling consumer credit risk and monitoring customer performance. Furthermore, the Gini index is known to be a transformation of the area under the ROC curve, which has been widely used for assessing performance of classification rules and facilitated prudent decision making in a variety of areas (e.g., Krzanowski \& Hand, 2009), including market segmentation (e.g., Benton \& Hand, 2002). Business analysts and managers, including those of sports teams, have used the Gini index to gain useful information about the market and consumer behaviour (e.g., Fort \& Quirk, 1995; Manasis et al., 2013; Manasis \&  Ntzoufras, 2014). Denneberg (1990) advocated the use of the Gini mean difference and the Gini index as safety loadings for insurance premiums.

Given all these and other diverse uses, we find a multitude of interpretations, mathematical expressions, and generalizations of the index in the literature. As noted by Ceriani \& Verme (2012), Corrado Gini himself had proposed no less than thirteen  formulations of his original index. Yitzhaki (1998, 2003), and Yitzhaki \& Schechtman (2013) have discussed a great variety of interpretations of the Gini index. Many monographs and handbooks have been written on measuring economic inequality where the Gini index and its various extensions and generalizations have  played prominent roles:
Amiel \&  Cowell (1999),
Atkinson \& Bourguignon (2000, 2015),
Atkinson \& Piketty (2007),
Banerjee \& Duflo (2011),
Champernowne  \&  Cowell (1998),
Cowell (2011),
Kakwani (1980a),
Lambert (2001),
Nyg{\aa}rd \& Sandstr\"om (1981),
Ostry et al. (2014),
Piketty (2014),
Sen (1997),
Silber  (1999), and
Yitzhaki \& Schechtman (2013), to name a few.

Given all this diversity, one naturally wonders if there is one underlying thread that unifies all these indices. The population Lorenz function, as well as its various distances from the hypotenuse of the right triangle where every Lorenz function falls into, have traditionally provided such a thread (cf., e.g, Zitikis, 2002, for a mathematical viewpoint of the problem). However, recent developments in the area of measuring economic inequality (cf., e.g., Palma, 2006; Zenga, 2007; Gastwirth, 2014) have highlighted the need for departure from the population mean, which is inherent in the definition of the Lorenz function as the benchmark or reference point for measuring economic inequality. The newly developed indices have deviated from the aforementioned unifying thread and thus initiated a fresh rethinking of the problem of measuring inequality.

Bennett \& Zitikis (2015) have recently made a step in this direction by  suggesting a way to bridge Harsanyi's (1953) and Rawls's (1971) conceptual frameworks via a spectrum of random societal positions. In this paper we make a further step by developing a mathematically rigorous approach for unifying and interpreting numerous classical and contemporary indices of economic inequality, as well as those of risk. Briefly, the approach we have developed is based on appropriately chosen
\begin{enumerate}[1)]
  \item
  societal references such as the population mean, median, or some population-tail based measures, and
  \item
  distributions of random personal positions or gambles that determine person's position on a certain population-based function.
\end{enumerate}

Certainly, the literature is permeated with discussions related to points 1) and 2). For example, the choice of appropriate reference measures and inherent relativity issues have been prominently discussed by Sen (1983, 1998), Duclos (2000), and many others. The relativity of economic inequality measurement has been present in virtually every work, due to the fact that the very basic underlying quantities such as the Lorenz function are relative quantities with respect to the population mean income. Furthermore, the construction of distributions that govern personal random positions on population-based functions have been explored within a variety of contexts, such as the dual or rank-dependent utility theory (Quiggin, 1982, 1993; Schmeidler, 1986, 1989; Yaari, 1987), other non-expected utility theories (e.g., Puppe, 1991; Machina, 1987, 2008; and references therein), distortion risk measures (Wang, 1995, 1998), and weighted insurance premium calculation principles (Furman \& Zitikis, 2008, 2009).

The rest of the paper is organized as follows. In Section \ref{gini} we revisit the classical Gini index and, in particular, express it in two ways --  absolute and relative -- within the framework of expected utility theory using appropriately chosen gambles and societal functions (i.e., Lorenz and Bonferroni). In Section \ref{mean-ref} we step aside from the Lorenz and Bonferroni functions and, crucially for this paper, suggest using a lower conditional expectation as the underlying societal function on which various personal gambles are played; yet, the reference measure still remains the mean income $\mu_F$. In Section \ref{gen-ref} we depart from the latter reference and introduce a general index that accommodates any population-based reference measure $\theta_F$. In Sections \ref{subs:DW} and \ref{subs:Wang}, we show how the  Donaldson-Weymark-Kakwani index and the Wang (or distortion) risk measure, as well as their generalizations, fall into the expected utility framework with collective mean-income references and appropriately chosen personal gambles. In Section \ref{ref-indiv} we argue for the need of incorporating personal preferences into reference measures, and in Section \ref{subs:relative} we demonstrate how this yields a new measure of risk that takes into account the relativity of large (e.g., insurance) losses with respect to smaller ones. The paper finishes with a general and highly-encompassing index of inequality and risk, defined and discussed in Section \ref{gen-risk}.

\section{\normalsize The classical Gini index revisited}
\label{gini}

Naturally, we begin our arguments with the classical index of Gini (1914). Let $X$ be an `income' random variable with non-negatively supported cdf $F(x)$ and finite mean value $\mu_F=\mathbf{E}[X]$. The Gini index, which we denote by $\mathrm{G}_F$, is usually interpreted as twice the area between the actual population Lorenz function (Lorenz, 1905; Pietra, 1915; Gastwirth, 1971)
\[
\mathrm{L}_F(p)={1 \over \mu_F } \int_0^p F^{-1}(t)dt 
\]
and the equalitarian Lorenz function $\mathrm{L}_E(p)=p$, $0\le p \le 1$, which is the hypotenuse of the right triangle that we have eluded to in the abstract. For parametric expressions of $\mathrm{L}_F(p)$, we refer to Gastwirth (1971), Kakwani \& Podder (1973), as well as to more complete and recent works by Kleiber \& Kotz (2003), Sarabia (2008), Sarabia et al. (2010), and references therein. Hence, the Gini index is
\begin{align}
\mathrm{G}_F
&=2\int_0^1 \big (\mathrm{L}_E(p)-\mathrm{L}_F(p) \big ) \,dp
\notag
\\
&=2\mathbf{E}\big [\mathrm{L}_E(\pi)-\mathrm{L}_F(\pi)\big ],
\label{gini-00}
\end{align}
where the gamble $\pi $ follows the uniform density on the unit interval $[0,1]$, that is, $f(p)=1$ for all $p\in [0,1]$. Intuitively, $\pi $ governs the person's position in terms of income percentiles, and we thus call it \textit{personal} gamble. In other words, barring the normalizing constant $2$, the Gini index $\mathrm{G}_F$ is the expected \textit{absolute}-deviation of the person's position $\pi $ on the actual Lorenz function $\mathrm{L}_F(p)$ from his/her position on the reference (equalitarian) Lorenz function $\mathrm{L}_E(p)$. Naturally, the position $\pi $ is random, and we have already seen in the case of the Gini index that it follows the uniform on $[0,1]$ distribution. This means that the person has equal chance of receiving any income among all the available incomes which are, in terms of percentiles, identified with the unit interval $[0,1]$.

\begin{figure}[h!]
\centering
\includegraphics[height=10cm, width=10cm]{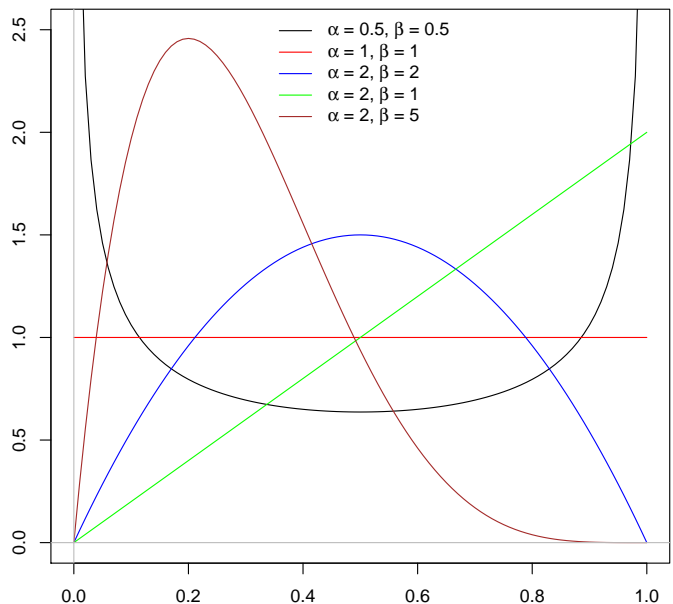}
\caption{Beta densities of gambles $\pi$ for various values of $\alpha$ and $\beta$.}
\label{fig:beta-density}
\end{figure}

In general, the personal gamble $\pi $  can follow various distributions on $[0,1]$, and we shall see a variety of examples throughout this paper. The choice of distribution of the gamble $\pi $ carries information about person's probable positions and is thus inevitably subjective, but many of the examples that we have encountered in the literature follow the beta distribution
\[
f_{\mathrm{Beta}}(p\mid \alpha,\beta )={p^{\alpha-1}(1-p)^{\beta-1}\over B(\alpha,\beta )} \quad \textrm{for} \quad 0< p < 1,
\]
which we have visualized in Figure \ref{fig:beta-density}.
We succinctly write $\pi\sim \mathrm{Beta}(\alpha,\beta )$ and so, for example, the Gini index (cf. equation (\ref{gini-00})) is based on $\pi\sim \mathrm{Beta}(1,1)$. For illuminating discussions and historical notes on the beta and other related distributions in the context of measuring economic inequality, we refer to Kleiber \& Kotz (2003). For a very general yet remarkably tractable beta-generated family of distributions that offers great modeling flexibility, we refer to Alexander et al. (2012).

Importantly for our following discussion, the Gini index $\mathrm{G}_F$ can also be viewed as the expected \textit{relative}-deviation of the person's position $\pi $ on the actual Lorenz function $\mathrm{L}_F(p)$ from his/her position on the reference Lorenz function $\mathrm{L}_E(p)$, as seen from the equations:
\begin{align}
\mathrm{G}_F
&=\int_0^1 \bigg (1-{\mathrm{L}_F(p)\over \mathrm{L}_E(p)} \bigg ) 2p\, dp
\notag
\\
&=\mathbf{E}\bigg [1-{\mathrm{L}_F(\pi)\over \mathrm{L}_E(\pi)}\bigg ],
\label{gini-01}
\end{align}
where $\pi \sim \mathrm{Beta}(2,1)$, which is a considerable change from the gamble $\pi\sim \mathrm{Beta}(1,1)$ used in the absolute-deviation based representation (\ref{gini-00}) of the Gini index. Note that the right-hand side of equation (\ref{gini-01}) can be succinctly written as $\mathbf{E}[\mathrm{B}_F(\pi )]$, where
\begin{equation}
\mathrm{B}_F(p)
=1-{\mathrm{L}_F(p)\over \mathrm{L}_E(p)}
=1-{\mathrm{L}_F(p)\over p}
\label{bonf-0}
\end{equation}
is the Bonferroni function of inequality (cf. Bonferroni, 1930), which is also known in the literature as the Gini function of inequality because it appeared in Gini (1914). For details on the Bonferroni function and the corresponding Bonferroni index, we refer to Tarsitano (1990) and references therein.

In addition to its role when studying income and poverty, the Bonferroni function $\mathrm{B}_F(p)$ has also found many uses in other fields such as reliability, demography, insurance, and medicine (cf.\, Giorgi \& Crescenzi, 2001; and references wherein). For detailed historical notes and references with explicit expressions for the Lorenz and Bonferroni functions, as well as for the Gini and Bonferroni indices, in the case of a great many parametric distributions, we refer to Giorgi \& Nadarajah (2010). The role of the Bonferroni function within the framework of $L$-functions for measuring economic inequality and actuarial risks can be found in Tarsitano (2004), and Greselin et al.\, (2009).

\section{\normalsize From equalitarian Lorenz to the mean reference}
\label{mean-ref}

Not only the classical Gini index but also a multitude of other indices of economic inequality can be viewed as some deviation measures (e.g., functional distances) between the actual and equalitarian Lorenz functions (cf., e.g., Zitikis, 2002). Note, however, that the actual Lorenz function $\mathrm{L}_{F}(p)$ itself is a relative measure that compares $p\times 100\%$ of lowest incomes with the population mean income $\mu_F$. This two-stage relativity -- first with respect to the equalitarian Lorenz function and then with the mean income -- warrants a serious rethinking of the inequality measurement.

Toward this end, we next rephrase the definition of the Gini index $\mathrm{G}_F$ by first rewriting the Bonferroni function $\mathrm{B}_F(p)$ as follows:
\begin{equation}
\mathrm{B}_F(p)=1-{\mathrm{LCE}_{F}(p)\over \mu_F},
\label{equate-0}
\end{equation}
where
\[
\mathrm{LCE}_{F}(p)={1\over p}\int_0^p F^{-1}(t)dt
\]
is the lower conditional expectation of $X$. Indeed, with a little mathematical caveat, $\mathrm{LCE}_{F}(p)$ is the conditional expectation $\mathbf{E}[X \mid X\le F^{-1}(p)]$, which is the mean income of those who are below the `poverty line' $F^{-1}(p)$. In summary, equation (\ref{gini-01}) becomes
\begin{equation}
\mathrm{G}_F=\mathbf{E}\bigg [1-{\mathrm{LCE}_{F}(\pi)\over \mu_F}\bigg ]
\label{gini-01b}
\end{equation}
with the gamble $\pi \sim \mathrm{Beta}(2,1)$.

Interestingly, if instead of the latter gamble we use $\pi \sim \mathrm{Beta}(1,1) $ on the right-hand side of equation (\ref{gini-01b}), then the expectation turns into the Bonferroni index
\begin{equation}
\mathrm{B}_F=\int_0^1 \bigg (1-{\mathrm{LCE}_{F}(\pi)\over \mu_F}\bigg ) dp.
\label{bonf-01b}
\end{equation}
For details on this index, we refer to Tarsitano (1990) and references therein. For a comparison of the two weighting schemes, that is, of the gambles $\pi$ employed in the Gini and Bonferroni cases, we refer to De Vergottini (1940). Implications of using the Bonferroni index on welfare measurement have been studied by, e.g., Benedetti (1986), Aaberge (2000), and Chakravarty (2007). Nyg{\aa}rd \& Sandstr\"{o}m (1981) give a wide-ranging discussion on the use of Bonferroni-type concepts in the measurement of economic inequality. Giorgi \& Crescenzi (2001), and Chakravarty \& Muliere (2004) propose poverty measures based on the fact that the Bonferroni index  exhibits greater sensitivity on lower levels of the income distribution than the Gini index. A general class of inequality measures inspired by the Bonferroni index has been explored by Imedio-Olmedo et al. (2011). Giorgi (1998) provides a list of Bonferroni's publications.

Equations (\ref{gini-01b}) and (\ref{bonf-01b}) suggest that the Gini and Bonferroni indices are members of the following general class of indices 
\begin{equation}
\mathcal{A}_F=\mathbf{E}[v(\mathrm{LCE}_F(\pi),\mu_F)],
\label{a-0}
\end{equation}
where $v(x,y)$ can be any function for which the expectation is well-defined and finite. In the case of the Gini and Bonferroni indices (cf., e.g., Greselin, 2014), we have  $v(x,y)=1-x/y$, which is the relative value of $x$ with respect to $y$; we call any function $v(x,y)$ used in expressions like (\ref{a-0}) a relative-value function throughout this paper. Hence, we can view the index $\mathcal{A}_F$ as the expected utility of being in the society whose income distribution is depicted by the lower conditional expectation $\mathrm{LCE}_{F}(p)$ and compared with the reference mean income $\mu_F$ using an appropriately chosen relative-value function $v(x,y)$.

The class of relative-value functions  $v(x,y)$ is very large, but in this paper we only deal with those that are of the form
\begin{equation}\label{ell}
v(x,y)=\ell (x/y)
\end{equation}
for some function $\ell (t)$. This form is natural because under the assumption of positive homogeneity on the function $v(x,y)$, which means that the equation $v(\lambda x,\lambda y)=v(x,y)$ holds for all $\lambda>0$, Euler's classical theorem says that we must have equation (\ref{ell}) for some function $\ell (t) $.

Hence, the Gini and Bonferroni indices produce the function $\ell (t)=1-t$. Another example of the function $\ell (t)$ arises from the $E$-Gini index of Chakravarty (1988):
\begin{align}
\mathrm{C}_{F,\alpha}
&= 2 \bigg (\int _{0}^{1}(t-\mathrm{L}_F(t))^{\alpha} dt\bigg ) ^{1/\alpha }
\notag
\\
&=  2 \bigg ( \int _{0}^{1} \bigg (1-{\mathrm{LCE}_{F}(\pi)\over \mu_F}\bigg )^{\alpha} t^{\alpha}dt\bigg ) ^{1/\alpha }
\notag
\\
&=  {2\over (\alpha +1)^{1/\alpha} }
\Big ( \mathbf{E}[v(\mathrm{LCE}_F(\pi),\mu_F)] \Big ) ^{1/\alpha },
\label{eq.12}
\end{align}
where the reference-value function is $v(x,y)=(1-x/y)^{\alpha}$, that is, $\ell (t)=(1-t)^{\alpha}$, and the gamble $\pi \sim \mathrm{Beta}(\alpha+1,1)$. Zitikis (2002) suggests using $(\alpha +1)^{1/\alpha}$ instead of 2 in the definition of the $E$-Gini index (see also Zitikis, 2003, for additional notes) in which case the right-hand side of equation (\ref{eq.12}) turns into the index
\[
\widetilde{\mathrm{C}}_{F,\alpha}=\Big (\mathbf{E}[v(\mathrm{LCE}_F(\pi),\mu_F)] \Big )^{1/\alpha }.
\]
In either case, note from the expressions of $\mathrm{C}_{F,\alpha}$ and $\widetilde{\mathrm{C}}_{F,\alpha}$ that it is sometimes useful to transform the index $\mathcal{A}_F$ by some function $w(x)$. We shall elaborate on this point in the next section.

Coming now back to the index $\mathcal{A}_F$, we note that with the generic relative-value function $v(x,y)=\ell (x/y)$, the index can be rewritten as $\mathbf{E}[\bar{\ell }(B_F(\pi))]$, where $\bar{\ell }(t)=\ell (1-t)$. Hence, we are dealing with the distorted Bonferroni function $\bar{\ell }(B_F(p))$, $0<p<1$, which is analogous to the distorted Lorenz function upon which Sordo et al. (2014) have built their far-reaching development of Aaberge's (2000) work. We do not pursue this research venue in the present paper because the Bonferroni function, just like that of Lorenz, incorporates a pre-specified reference measure, which is the mean income $\mu_F$. In what follows, we shall argue in favour of more flexibility when choosing reference measures, which may even incorporate personal preferences in addition to those of the entire population.

\section{\normalsize From the mean to generic societal references}
\label{gen-ref}

We are now in the position to extend the index $\mathcal{A}_F$ to arbitrary references, which we denote by $\theta_F$. Namely, let
\[
\mathcal{B}_F=w\Big (\mathbf{E}[v(\mathrm{LCE}_F(\pi),\theta_F)]\Big ),
\]
where $w(x)$ is a `normalizing' function whose main role is to fit the index into the unit interval $[0,1]$, with value $0$ meaning `perfect equality' and $1$ `extreme inequality.' The flexibility to manipulate with references is important due to a variety of reasons. For example, the increasing skewness of populations could make  the use of the mean $\mu_F$ questionable, and this has been noted by Gastwirth (2014) who, in his research on the changing income inequality in the U.S. and Sweden, has suggested replacing the mean $\mu_F$ by the median $m_F=F^{-1}(0.5)$.

Another example of $\theta_F$ that differs from $\mu_F$ is provided by the  Palma index (cf., Cobham \& Sumner, 2103a,b, 2014). Namely, let $\theta_F$ be the average of the top 10\% of the population, that is, $\theta_F=\frac{1}{0.1}\int_{0.9}^1 F^{-1}(t)dt$. Furthermore, let the normalizing function be $w(x)=x$, the relative-value function $v(x,y)=y/x $, and the (deterministic) gamble $\pi =0.4$. Under these specifications, the index $\mathcal{B}_F$ reduces to the Palma index of economic inequality:
\[
\mathrm{P}_F^{40,90}
={\frac{1}{0.1}\int_{0.9}^1 F^{-1}(t)dt \over \frac{1}{0.4}\int_0^{0.4} F^{-1}(t)dt} .
\]

Instead of the underlying random variable (e.g., income) $X$, the researcher might be primarily interested in its transformation (e.g., utility of income) $u(X)$. To tackle this situation, we first incorporate the transformed incomes into our framework by extending the definition of the lower conditional expectation as follows:
\[
\mathrm{LCE}_{F,u}(p)={1\over p}\int_0^p u(F^{-1}(t))dt.
\]
Note that $\mathrm{LCE}_{F,u}(1)=\mathbf{E}[u(X)]$, which we can view as the expected utility of $X$. We have arrived at the extension
\[
\mathcal{C}_F=w\Big (\mathbf{E}[v(\mathrm{LCE}_{F,u}(\pi),\theta_F)] \Big)
\]
of the index $\mathcal{B}_F$.

The Atkinson (1970) index, which we denote by $\mathrm{A}_{F,\gamma}$, is a special case of $\mathcal{C}_F$. Namely, let the utility function be $u(x)=x^{\gamma }$ for some $\gamma \in (0,1)$. Furthermore, let the (deterministic) gamble be $\pi=1$, the reference $\theta_F=u(\mu_F)$, and the relative-value function $v(x,y)=1-x/y$. Under these specifications, the index $\mathcal{C}_F$ turns into $1-\mathbf{E}[X^{\gamma }] / \mu_F^{\gamma }$, which after the transformation with the function $w(x)=1-(1-x)^{1/\gamma }$ becomes the Atkinson index
\[
\mathrm{A}_{F,\gamma}=1-{(\mathbf{E}[X^{\gamma }])^{1/\gamma } \over \mu_F}.
\]
This index has been highly influential in measuring economic inequality (cf., e.g., Cowell, 2011, and references therein) and inspired a variety of extensions and generalization of the Gini index; we shall say more on this in the next section. Furthermore, Mimoto \& Zitikis (2008) have found the Atkinson index useful in developing a statistical inference theory for testing exponentiality, which has been a prominent problem in life-time analysis and particularly in reliability engineering.

\section{\normalsize Donaldson-Weymark-Kakwani index}
\label{subs:DW}

The Donaldson-Weymark-Kakwani index (Donaldson \& Weymark, 1980, 1983; Kakwani, 1980a,b; Weymark, 1981)
\[
\mathrm{DWK}_{F,\alpha}
=\alpha(\alpha-1)\int_0^1 (1-p)^{\alpha-2} (p-\mathrm{L}_F(p))dp,
\]
which is also known as the $S$-Gini index, has arisen following Atkinson's (1970) criticism of the Gini index $\mathrm{G}_F$ for not being able to take into account social preferences. Via the parameter $\alpha>1$, the index $\mathrm{DWK}_{F,\alpha}$ can reflect different social preferences, with the classical Gini index arising by setting $\alpha=2$. We note in this regard that a justification for a family of indices to be based on the theory of relative deprivation has been provided by Yitzhaki (1979, 1982).

\begin{figure}[h!]
\centering
\includegraphics[height=10cm, width=10cm]{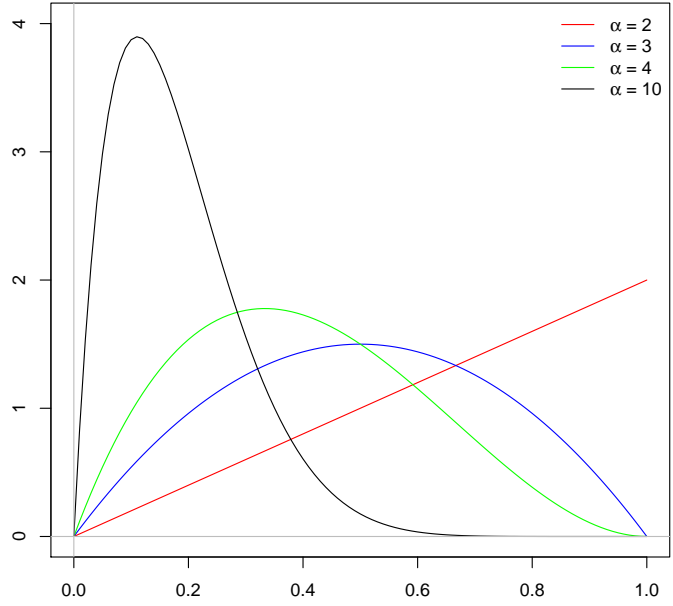}
\caption{The density of $\pi_{\alpha }$ for various values of $\alpha$.}
\label{fig:DW-density}
\end{figure}

Just like the Gini index $\mathrm{G}_F$, the index  $\mathrm{DWK}_{F,\alpha}$ can also be placed within the framework of expected relative value. Indeed, using equations (\ref{bonf-0}) and (\ref{equate-0}), we have
\begin{align}
\mathrm{DWK}_{F,\alpha}
&=\int_0^1
\bigg (1-{\mathrm{L}_F(p)\over p}\bigg )f_{\mathrm{Beta}}(p\mid 2,\alpha-1)dp
\notag
\\
&=\int_0^1
\bigg (1-{\mathrm{LCE}_{F}(p)\over \mu_F}\bigg )f_{\mathrm{Beta}}(p\mid 2,\alpha-1)dp
\notag
\\
&=\mathbf{E}[v(\mathrm{LCE}_{F}(\pi_{\alpha }),\mu_F )]
\label{dw-11a}
\end{align}
with the relative-value function  $v(x,y)=1-x/y$ and the gamble  $\pi_{\alpha } $ that follows the distribution $ \mathrm{Beta}(2,\alpha-1)$, whose density we have visualized in Figure \ref{fig:DW-density}.

We next introduce a more flexible index than $\mathrm{DWK}_{F,\alpha}$ that allows us to employ more general gambles than $\pi_{\alpha }$. For this, we first introduce a class of generating functions:
\begin{description}
\item[(H)] Let $h:[0,1] \to [0,1]$ be any twice differentiable and convex function (i.e., $h''(t) \geq 0$ for all $t\in (0,1)$) that satisfies the boundary conditions $h(0)=0 $ and $ h(1)=1$, and such that
$h'(0) \neq 1 $.
\end{description}
Let $\pi_h $ denote the gamble whose density $f(t)$ is given by the formula
\begin{equation}\label{choice-dw}
f(t)={t \, h''(1-t) \over 1-h'(0)}
\end{equation}
for all $t\in (0,1)$, and  $f(t)=0$ elsewhere. With the relative-value function $v(x,y)=1-x/y$, we have (details in Appendix \ref{appendix})
\begin{align}
\mathrm{DWK}_{F,h}
&:=\mathbf{E}[v(\mathrm{LCE}_F(\pi_h),\mu_F)]
\notag
\\
&= {1\over 1- h'(0)} \bigg ( 1- {1\over \mu_F }\int_0^1  F^{-1}(t) h'(1-t) \, dt \bigg )
\notag
\\
&= {1\over 1- h'(0)} \bigg ( 1- {1\over \mu_F }\int_0^\infty  h(1-F(x)) \, dx\bigg ).
\label{eq-dw-0b}
\end{align}

To illustrate, we choose the function $h(t)=t^{\alpha}$ with any $\alpha >1$, in which case the gamble $\pi_h $ follows the density  $\alpha (\alpha-1)p(1-p)^{\alpha-2}$; that is, $\pi_h \sim \mathrm{Beta}(2,\alpha-1)$, which means that $\pi_h$ has the same distribution as the earlier noted gamble $\pi_{\alpha }$. Consequently, $\mathrm{DWK}_{F,h}$ reduces to $\mathrm{DWK}_{F,\alpha}$, and thus  equations (\ref{eq-dw-0b}) reduce to the following expressions of the Donaldson-Weymark-Kakwani index:
\begin{align}
\mathrm{DWK}_{F,\alpha}
&=1 - \frac{\alpha}{\mu_F}\int_0^1  F^{-1}(t) (1-t)^{\alpha-1} dt
\notag
\\
&=1- \frac{1}{\mu_F}\int_0^{\infty } (1-F(x))^\alpha dx
\label{dw-alternative-2}
\end{align}
(cf.\, Donaldson \& Weymark, 1980, 1983; Yitzhaki, 1983; Muliere \& Scarsini, 1989).

\section{\normalsize Wang risk measure}
\label{subs:Wang}

The index $\mathrm{DWK}_{F,h}$ is based on gambles generated by \textit{convex} functions $h$. A similar index but based on \textit{concave} generating functions $g$ is called the Wang (or distortion) risk measure,
which has been used in actuarial science and financial mathematics for measuring risks. In detail, the risk measure is defined by the formula
\[
\mathrm{W}_{F,g}=\int_0^\infty  g(1-F(x)) \, dx,
\]
where $g:[0,1] \to [0,1]$ is a distortion function, meaning that it is non-decreasing and satisfies the boundary conditions $g(0)=0 $ and $ g(1)=1$.

Hence, unlike in the previous section, we now work with concave distortion functions, denoted by $g$, under which the risk measure $\mathrm{W}_{F,g}$ is coherent (Wang et al., 1997; Wang \& Young, 1998; Wirch \& Hardy, 1999; see Artzner et al., 1999, for a general discussion). A classical example of such a distortion function is $g(t)=t^\alpha$ for any $\alpha \in (0,1)$, in which case the Wang risk measure $\mathrm{W}_{F,g}$ reduces to the proportional-hazards-transform risk measure (Wang, 1995)
\[
\mathrm{PHT}_{F,\alpha}=\int_0^\infty(1-F(x))^{\alpha} \, dx.
\]
For more information on concave vs convex distortion functions in the context of measuring risks, their variability and orderings, we refer to Sordo \& Su\'{a}rez-Llorens (2011), Giovagnoli \& Wynn (2012), and references therein.

We shall next show that the Wang risk measure $\mathrm{W}_{F,g}$ can be placed within the framework of expected relative value. When compared with the index $\mathrm{DWK}_{F,\alpha}$, there are two major changes: First, the function of interest is now the upper conditional expectation:
\[
\mathrm{UCE}_F(p)={1\over 1-p}\int_p^1 F^{-1}(t)dt .
\]
(Note that when $p=0$, then $\mathrm{UCE}_F(p)$ is equal to the mean $\mu_F$.) Second, the function $g$ that generates the distribution of the random position is concave; specifically, we introduce the following class of generating functions:
\begin{description}
\item[(G)] Let $g:[0,1] \to [0,1]$ be twice differentiable and concave function (i.e., $g''(t) \leq 0$ for all $t\in (0,1)$) that satisfies the boundary conditions $g(0)=0 $ and $ g(1)=1$, and such that $g'(1) \neq 1 $.
\end{description}
Any such function $g$ generates the density $f(t)$ of the gamble $\pi_g $ given by the formula
\begin{equation}\label{choice-wang}
f(t)={-(1-t) g''(1-t)\over 1-g'(1)}
\end{equation}
for all $t\in (0,1)$, and $f(t)=0$ elsewhere. With the relative-value function $v(x,y)=x/y-1$, we have (details in Appendix \ref{appendix})
\begin{align}
\mathbf{E}[v(\mathrm{UCE}_F(\pi_g),\mu_F)]
&= {1\over 1- g'(1)} \bigg ({1\over \mu_F }\int_0^1  F^{-1}(t) g'(1-t) \, dt -1\bigg )
\notag
\\
&= {1\over 1- g'(1)} \bigg ( {1\over \mu_F }\int_0^\infty  g(1-F(x)) \, dx -1 \bigg ).
\label{eq-w-0b}
\end{align}
Consequently, the Wang risk measure $\mathrm{W}_{F,g}$ can be expressed in terms of the expected relative value $\mathbf{E}[v(\mathrm{UCE}_F(\pi_g),\mu_F)]$ as follows:
\begin{equation}\label{wang-10}
\mathrm{W}_{F,g}=\mu_F\Big ( \mathbf{E}[v(\mathrm{UCE}_F(\pi_g),\mu_F)]\big(1- g'(1)\big )+1\Big ).
\end{equation}

\begin{figure}[h!]
\centering
\includegraphics[height=10cm, width=10cm]{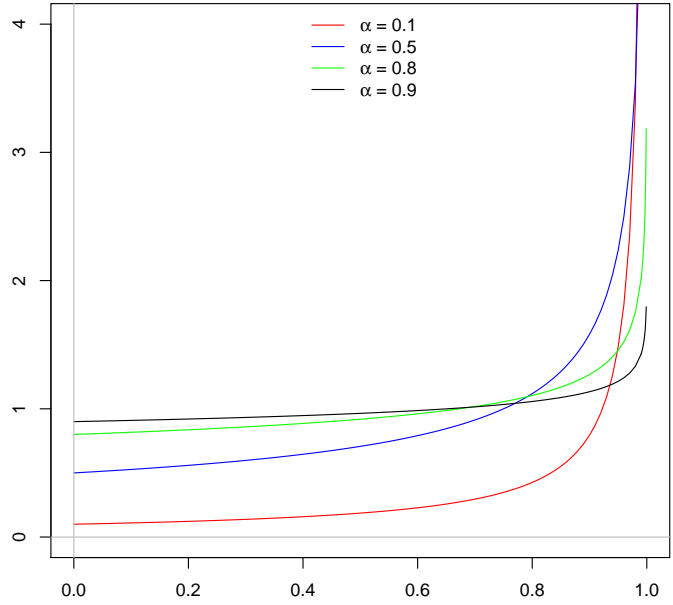}
\caption{The density of $\pi_g$ when $g(t)=t^{\alpha}$ for various values of $\alpha$.}
\label{fig:Wang-density}
\end{figure}

When the generating function is $g(t)=t^\alpha$ for any $\alpha \in (0,1)$, then the gamble $\pi_g $ follows the distribution $\mathrm{Beta}(1,\alpha)$ whose density function $\alpha(1-p)^{\alpha-1}$ is depicted in Figure \ref{fig:Wang-density}.
From equation (\ref{eq-w-0b}) we have
\begin{equation}\label{eq-w-00}
\mathbf{E}[v(\mathrm{UCE}_F({_{\alpha}}\pi),\mu_F)]
={1\over 1-\alpha } \bigg ( \frac{1}{\mu_F}\int_0^\infty(1-F(x))^{\alpha} \, dx -1 \bigg ).
\end{equation}
Consequently, we have the following expression for the proportional-hazards-transform risk measure:
\[
\mathrm{PHT}_{F,\alpha}=\mu_F\Big ( \mathbf{E}[v(\mathrm{UCE}_F(\pi_g),\mu_F)]\big (1- \alpha \big )+1\Big ).
\]

\section{\normalsize From collective to individual references}
\label{ref-indiv}

So far, we have worked with \textit{collective} references; they do not depend on the outcome of any personal gamble and thus apply to every member of the society. Such references may not, however, be always desirable or justifiable. For example, given the outcome $0.4$ of the gamble $\pi $, meaning that the person is considered to be among the $p=40\% $ of the lowest income earners, the person may wish to compare his/her situation to that as if he/she were among $1-p=60\% $ of the highest income earners. In such situations we are dealing with \textit{individual} references; their values may depend on outcomes of the personal gamble $\pi $.

Hence, for example, the mean $\mu_F$ and the median $m_F=F^{-1}(0.5)$ are collective references, but $\theta_F=F^{-1}(\pi)$ is an individual reference because its value depends on the outcome of $\pi $. But would the quantile $F^{-1}(\pi)$ be a good reference? There are at least two major reasons against the use of the quantile, which is known in the risk literature as the value-at-risk:
\begin{enumerate}[1)]
  \item
  The quantile $F^{-1}(\pi)$ is not robust with respect to realized values of the random gamble $\pi$ in the sense that the quantile may change drastically even for very small changes of the realized values of $\pi$.
  \item
  For a given realized value of $\pi$, the quantile $F^{-1}(\pi)$ is not informative about the values of $F^{-1}(p)$ for $p>\pi$. Indeed, for a given realized-value of $\pi$, we may see the same value of $F^{-1}(\pi)$ irrespective of whether the cdf $F$ is heavy- or light-tailed.
\end{enumerate}

These are serious issues when constructing sound measures of economic inequality and risk. In the risk literature (cf., e.g., McNeil et al., 2005; Meucci, 2007; Pflug \& R\"{o}misch, 2007; Cruz, 2009; Sandstr\"{o}m, 2010; Cannata \& Quagliariello, 2011; and references therein) the problem with quantiles has been overcome by using the \textit{tail}-value-at-risk,  which is the upper conditional expectation whose definition was given in the previous section. For example, adopting $\mathrm{UCE}_F(\pi)$ as our (individual) reference $\theta_F$ and using the normalizing function $w(x)=x$, the earlier introduced index $\mathcal{B}_F$ turns into the Zenga (2007) index
\begin{align}
\mathrm{Z}_F
&=\int_0^1\bigg(1-{\mathrm{LCE}_F(p)\over \mathrm{UCE}_F(p)}\,\bigg )\,dp
\notag
\\
&=\mathbf{E}[v(\mathrm{LCE}_F(\pi),\mathrm{UCE}_F(\pi))]
\label{zenga-0}
\end{align}
with the relative-value function $v(x,y)=1-x/y$ and the gamble $\pi\sim \mathrm{Beta}(1,1)$. Hence, the Zenga index $\mathrm{Z}_F$ is the average  with respect to all percentiles $p\in (0,1)$ of the relative deviations of the mean income of the poor (i.e., those whose incomes are below the poverty line $F^{-1}(p)$) from the corresponding mean income of the rich, that is, of those whose incomes are above the poverty line $F^{-1}(p)$. We refer to Greselin et al.\, (2013) for a more detailed discussion of the relative nature of the Gini and Zenga indices, and their comparison.

\section{\normalsize Relative measure of risk}
\label{subs:relative}

Many risk measures that have appeared in the literature are designed to measure \textit{absolute} heaviness of the right-hand tail of the underlying loss-distribution. Suppose now that we wish to measure the severity of large (e.g., insurance) losses \textit{relative} to small ones. Notice that this problem is very similar to that tackled by Zenga (2007) in the context of economic inequality. Hence, following the same path but now using the relative-value function $v(x,y)=y/x-1$ and generic gamble $\pi $, we arrive at the relative measure of risk
\begin{equation}
\mathrm{R}_F
=\mathbf{E}[v(\mathrm{LCE}_F(\pi),\mathrm{UCE}_F(\pi))].
\label{r-0}
\end{equation}
Of course, given the above relative-value function, we can rewrite this risk measure in the spirit of expected utility
\begin{equation}
\mathrm{R}_F =\mathbf{E} [ \mathrm{R}_F(\pi)],
\label{r-1}
\end{equation}
where the role of utility function is played by the risk function
\[
\mathrm{R}_F(p)= {\mathrm{UCE}_F(p)\over \mathrm{LCE}_F(p)} \,-1 .
\]
In what follows we shall explore properties of this risk measure. We shall use the notation $\mathrm{R}_X$ instead of $\mathrm{R}_F$ when a need arises to emphasize the dependence of the risk measure on the random variable $X$ itself.

\begin{property}
We have the following statements:
\begin{enumerate}[\rm (i)]
\item\label{prop-i}
If the risk $X$ is constant, that is, $X=d$ for some constant $d>0$, then $\mathrm{R}_X=0$.
\item\label{prop-ii}
Multiplying the risk $X$ by any constant $d>0$ does not change the relative measure of risk, that is, $\mathrm{R}_{dX}=\mathrm{R}_X$.
\item \label{prop-iii}
Adding any constant $d>0$ to the risk $X$ decreases the relative measure of risk, that is,
$\mathrm{R}_{X+d} \le \mathrm{R}_X$.
\end{enumerate}
\label{prop-0}
\end{property}

The proof of Property \ref{prop-0} is technical and we have thus relegated it to Appendix \ref{appendix}. We shall next comment on the meaning of Property \ref{prop-0}. First, given that we are dealing with a \textit{relative} measure of risk, points (\ref{prop-i}) and (\ref{prop-ii}) are self-explanatory. As to point (\ref{prop-iii}), it says that lifting up the risk by any positive constant decreases its riskiness. This is natural because lifting up diminishes the relative variability of the risk. This, in turn, suggests that the ordering of relative risk measures should be done, for example, in terms of the Lorenz ordering, which is the most used tool for comparing the variability of economic-size distributions. This leads us to the following property:

\begin{property}\label{prop:Lorenz}
If risks $X$ and $Y$ follow the Lorenz ordering $X \leq_{\mathrm{L}}Y$, then $\mathrm{R}_{X}\le \mathrm{R}_Y$.
\end{property}

\begin{proof}
To verify Property \ref{prop:Lorenz}, we first recall (Arnold, 1987; see also Aaberge, 2000) that the Lorenz ordering $X \leq_{\mathrm{L}}Y$ means the bound $\mathrm{L}_X(p) \ge \mathrm{L}_Y(p)$ for all $p \in [0,1]$. Since
\begin{align*}
\mathrm{R}_X(p)
&= \frac{1-\mathrm{L}_X(p)}{\mathrm{L}_X(p)} \,\, \frac{p}{1-p}-1
\\
&=\frac{p}{(1-p)\mathrm{L}_X(p)}-\frac{p}{1-p}-1 ,
\end{align*}
the Lorenz ordering $X \leq_{\mathrm{L}}Y$ is therefore equivalent to the $R$-ordering $X \leq_{\mathrm{R}}Y$, which means $\mathrm{R}_{X}(p)\le \mathrm{R}_Y(p)$ for all $p \in (0,1)$. The latter bound and equation (\ref{r-1}) conclude the verification of Property \ref{prop:Lorenz}. Note also that with the notion of $R$-ordering, we can rephrase Property \ref{prop:Lorenz} as follows: if $X \leq_{\mathrm{R}}Y$, then $\mathrm{R}_{X}\le \mathrm{R}_Y$. For detailed treatments of various notions of stochastic orders, we refer to Shaked \& Shantikumar (2007), and Li \& Li  (2013).
\end{proof}

We next discuss another closely related ordering property, called the Pigou-Dalton principle of transfers. In the context of economic inequality, the principle says that progressive (i.e., from rich to poor) rank-order and mean-preserving transfers should decrease the value of inequality measures. Hence, in the complementary risk-framework, the transfers are risk-decreasing transformations. Formally (cf., Vergnaud, 1997), $X$ is less risk-unequal than $Y$ in the Pigou-Dalton sense, denoted by $X\le_{\mathrm{PD}}Y$, if and only if $\mu_X=\mu_Y$ and $X \leq_{\mathrm{L}}Y$. (Hence, $X\le_{\mathrm{PD}}Y$ is sometimes denoted by $X \leq_{\mathrm{L},\,=}Y$.) The following property is now obvious.

\begin{property}
If a Pigou-Dalton risk-increasing transfer turns risk $X$ into $Y$ so that $X\le_{\mathrm{PD}}Y$, then $\mathrm{R}_{X} \le \mathrm{R}_Y$.
\end{property}

To have an idea of how the Pigou-Dalton transfers act, we recall (e.g., Shaked \& Shantikumar, 2007; Li \& Li, 2013) that given $X$ and $Y$ with densities $f_X$ and $f_Y$ respectively, and assuming that their means are equal, if the sign of the difference $f_X-f_Y$ changes twice according to the pattern  $(+,-,+)$, then $X \leq_{\mathrm{L}}Y$. Examples of parametric distributions with such pdf's can be found in, e.g., Kleiber \& Kotz~(2003).

\begin{figure}[h!]
\centering
\includegraphics[height=10cm, width=10cm]{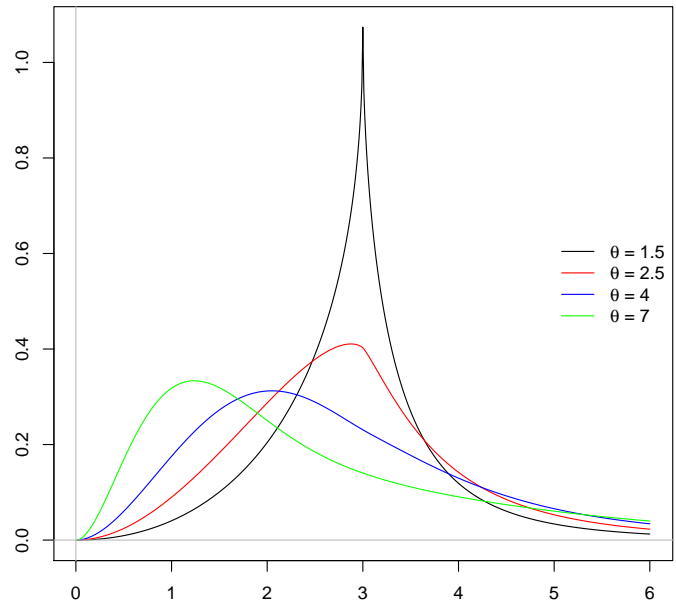}
\caption{$\mathrm{Zenga} (2, 2,\theta)$ density for various values of $\theta$.}
\label{fig:PD-property}
\end{figure}

In what follows, we shall discuss an example based on Zenga's (2010) distribution that has demonstrated remarkably good performance in terms of goodness-of-fit on a number of real income data sets. Namely, it is a very flexible three-parameter distribution with Pareto-type right-hand tail and whose density is
\[
f_{\mathrm{Zenga}}(x\mid \mu,\alpha,\theta)=\left\{
\begin{array}{ll}
\displaystyle
\frac{1}{2\mu \,\mathrm{Beta}(\alpha,\theta)}\bigg (\frac{x}{\mu}\bigg )^{-1.5} \int_{0}^{x/\mu}t^{\alpha+0.5-1}(1-t)^{\theta-2}dt, & x<\mu ,
\\
\\
\displaystyle
\frac{1}{2\mu \,\mathrm{Beta}(\alpha,\theta)}\bigg (\frac{\mu}{x} \bigg )^{1.5} \int_{0}^{\mu/x}t^{\alpha+0.5-1}(1-t)^{\theta-2}dt, & x\ge \mu ,
\end{array}\right.
\]
where $\mu $ is the scale parameter, which also happens to be the mean of the model, and $\theta$ and $\alpha$ are two shape parameters that affect, respectively, the center and the tails of the distribution. In Figure \ref{fig:PD-property}
we have illustrated the Zenga density. For further details on this  distribution and its uses, we refer to Zenga (2010), Zenga et al.~(2011, 2012), and Arcagni \& Zenga (2013). To see the effects of the Pigou-Dalton transfers in the case of the Zenga distribution, the following theorem is particularly useful.

\begin{theorem}[Arcagni \& Porro, 2013]
Assume $X \sim \mathrm{Zenga} (\mu_X, \alpha_X,\theta_X)$ and $Y \sim \mathrm{Zenga} (\mu_Y, \alpha_Y,\theta_Y)$, where all the parameters are positive. When $\alpha_X \geq \alpha_Y$ and $\theta_X \leq \theta_Y$, then  $X \leq_{\mathrm{PD}} Y$.
\end{theorem}

\section{\normalsize A general index of inequality and risk}
\label{gen-risk}

Remarkably, the right-hand sides of equations (\ref{zenga-0}) and (\ref{r-0}), which are identical barring their different relative-value functions $v(x,y)$, can -- with a little tweak -- turn into a very general measure of inequality: $\mathbf{E}[v(\mathrm{LCE}_F(\pi),\mathrm{UCE}_F(\pi^*))]$, where $\pi $ and $ \pi^*$ are two gambles, which could be dependent or independent, degenerate or not. Obviously, when $\pi =\pi^*$, then we get either the Zenga index of economic inequality or the relative measure of risk, depending on the relative-value function. Furthermore, if $\pi^*=0$, then we have $\mathrm{UCE}_F(\pi)=\mu_F$ and thus $\mathbf{E}[v(\mathrm{LCE}_F(\pi),\mu_F)]$, which is the Bonferroni index $\mathrm{B}_F$. By appropriately choosing relative-value functions and personal gambles, we can reproduce a number of other measures of economic inequality and risk, but the Chakravarty and Atkinson indices require some little extension:
\begin{equation}
\mathcal{E}_F=w\Big( \mathbf{E}\big [v\big (\mathrm{LCE}_{F,u}(\pi),
\mathrm{UCE}_{F,u^*}(\pi^*)\big )\big ]\Big),
\label{J}
\end{equation}
where $u$ and $u^*$ are two utility functions, and
\[
\mathrm{UCE}_{F,u^*}(p)={1\over 1-p}\int_p^1 u^*(F^{-1}(t))dt.
\]
(Note that $\mathrm{UCE}_{F,u^*}(0)=\mathbf{E}[u^*(X)]$.) All the examples that we have mentioned in this paper, and also many other ones that appear in the literature, are special cases of the just introduced index $\mathcal{E}_F$ (e.g., Table \ref{table-index}).
\begin{table}[h!]
 \medskip
 \centering
\rotatebox{90}{
\begin{tabular}{l|cccccc}
  \hline\hline
   & $\pi $ & $\pi^*$ & $w(x)$ & $v(x,y)$ & $u(x)$ & $u^*(x)$\\
  \hline
Atkinson $\mathrm{A}_{F,\alpha}$ & $1$ & $0$ & $1-(1-x)^{1/\gamma }$ & $1-x/y$ & $x^{\gamma }$ & $x$ \\
Bonferroni $\mathrm{B}_F$ & $\mathrm{Beta}(1,1)$ & $0$ & $x$ & $1-x/y$ & $x$ & $x$ \\
Chakravarty $\mathrm{C}_{F,\alpha}$ & $\mathrm{Beta}(\alpha+1,1)$ & $0$ & $2(\alpha +1)^{-1/\alpha}x^{1/\alpha }$ & $(1-x/y)^{\alpha}$ & $x$ & $x$ \\
Inequality index $\widetilde{\mathrm{C}}_{F,\alpha}$ & $\mathrm{Beta}(\alpha+1,1)$ & $0$ & $x^{1/\alpha }$ & $(1-x/y)^{\alpha}$ & $x$ & $x$ \\
Donaldson-Weymark-Kakwani $\mathrm{DWK}_{F,\alpha} $ & $\mathrm{Beta}(2,\alpha-1)$ & $0$ & $x$ & $1-x/y$ & $x$ & $x$ \\
Inequality index $\mathrm{DWK}_{F,h} $ & $\pi_{h}$ & $0$ & $x$ & $1-x/y$ & $x$ & $x$ \\
Gini $\mathrm{G}_F$ & $\mathrm{Beta}(2,1)$ & $0$ & $x$ & $1-x/y$ & $x$ & $x$ \\
Palma $\mathrm{P}_F^{40,90}$ & $0.4$ & $0.9$ & $x$ & $y/x$ & $x$ & $x$ \\
Risk measure $\mathrm{R}_F$ & Any & $\pi^*=\pi$ & $x$ & $y/x-1$ & $x$ & $x$ \\
Wang $\mathrm{W}_{F,g}$ &$1$ & $\pi_g$ & $\mu_F(x(1- g'(1))+1)$ & $y/x-1$ & $x$ & $x$ \\
Proportional hazards transform $\mathrm{PHT}_{F,\alpha}$ &$1$ & $\mathrm{Beta}(1,\alpha)$ & $\mu_F(x(1- \alpha)+1)$ & $y/x-1$ & $x$ & $x$ \\
Zenga $\mathrm{Z}_F$ &$\mathrm{Beta}(1,1)$ & $\pi^*=\pi$ & $x$ & $1-x/y$ & $x$ & $x$ \\
  \hline
\end{tabular}}
  \caption{Special cases of index (\ref{J}).}
  \label{table-index}
\end{table}

We conclude this section with the note that in the examples throughout this paper, the gambles $\pi$ and $\pi^*$ have been such that either they are identical (i.e., $\pi=\pi^*$) or one of them is degenerate (e.g., $\pi=1$ or $\pi^*=0$). There is no reason why this should always be the case: the two gambles can be dependent but not necessarily identical or degenerate. This suggests that, in general, modeling probability distributions of the pair $(\pi,\pi^*)$ can be conveniently achieved by, for example, specifying  1) marginal distributions of the gambles $\pi$ and $\pi^*$, and also 2) a dependence structure between the two gambles using an appropriately chosen copula. For methodological and applications-driven developments related to copulas, we refer to Nelsen (2006), Jaworski et al.~(2010), Jaworski et al.~(2013), and references therein.

\section*{\normalsize Acknowledgments}

We started working on this paper on a beautiful sunny day in December 2014, in the courtyard of Sant'Anna School of Advanced Studies in Pisa, Italy. The second author is grateful to the University of Milano-Bicocca for making his most inspiring scientific visit in Milan and Pisa possible.

%

\bigskip
\begin{center}
\textsc{References}
\end{center}
\medskip
\def\hang{\hangindent=\parindent\noindent}

\hang
\textsc{Aaberge, R.} (2000)
Characterizations of Lorenz curves and income distributions.
\textit{Social Choice and Welfare}, \textbf{17}, 639--653.

\hang
\textsc{Alexander, C., Cordeiro, G. M., Ortega, E. M. M. \& Sarabia, J. M.} (2012)
Generalized beta-generated distributions.
\textit{Computational Statistics and Data Analysis}, \textbf{56}, 1880--1897.

\hang
\textsc{Amiel, Y. \& Cowell, F. A.}  (1999)
\textit{Thinking About Inequality.}
Cambridge: Cambridge University Press.

\hang
\textsc{Arcagni, A. \& Porro, F.} (2013)
On the parameters of Zenga distribution.
\textit{Statistical Methods and Applications}, \textbf{22}, 285--303.

\hang
\textsc{Arcagni, A. \& Zenga, M.} (2013)
Application of Zenga's distribution to a panel survey on household incomes of European Member States.
\textit{Statistica \& Applicazioni},  \textbf{11},  79--102.

\hang
\textsc{Artzner, P., Delbaen F., Eber, J. M. \& Heath, D.} (1999) Coherent measures of risk. \textit{Mathematical Finance}, \textbf{9}, 203--228.

\hang
\textsc{Arnold, B. C.} (1987)
\textit{Majorization and the Lorenz Order: A Brief Introduction.}
New York: Springer.

\hang
\textsc{Atkinson, A. B.} (1970)
On the measurement of inequality.
\textit{Journal of Economic Theory}, \textbf{2}, 244--263.

\hang
\textsc{Atkinson, A. B. \& Bourguignon, F.}  (eds) (2000)
\textit{Handbook of Income Distribution}, vol.~1.
Amsterdam: Elsevier.

\hang
\textsc{Atkinson, A. B. \& Bourguignon, F.} (eds) (2015)
\textit{Handbook of Income Distribution}, vol.~2.
Amsterdam: Elsevier.

\hang
\textsc{Atkinson, A. B. \& Piketty, T.}  (eds) (2007) 
\textit{Top Incomes Over the Twentieth Century: A Contrast Between Continental European and English-speaking Countries.} Oxford: Oxford University Press.

\hang
\textsc{Banerjee, A. V. \& Duflo, E.} (2011)
\textit{Poor Economics: A Radical Rethinking of the Way to Fight Global Poverty.}
New York: Public Affairs.

\hang
\textsc{Bennett, C. J. \&  Zitikis, R.} (2015)
Ignorance, lotteries, and measures of economic inequality.
\textit{Journal of Economic Inequality}, \textbf{13}, 309--316.

\hang
\textsc{Benedetti, C.} (1986)
Sulla interpretazione benesseriale di noti indici di concentrazione e di altri.
\textit{Metron}, \textbf{44}, 421--429.

\hang
\textsc{Benton, T. C. \& Hand, D. J.} (2002)
Segmentation into predictable classes.
\textit{IMA Journal of Management Mathematics}, \textbf{13}, 245--259.

\hang
\textsc{Bonferroni, C. E.} (1930)
\textit{Elementi di Statistica Generale}. Firenze: Libreria Seeber.

\hang
\textsc{Cannata F. \&  Quagliariello M.} (2011)
\textit{Basel III and Beyond.}  London: Risk Books.

\hang
\textsc{Ceriani, L. \& Verme, P.} (2012)
The origins of the Gini index: extracts from Variabilit{\`a} e Mutabilit{\`a} (1912) by Corrado Gini.
\textit{Journal of Economic Inequality}, \textbf{10}, 421--443.

\hang
\textsc{Chakravarty, S. R.} (1988)
Extended Gini indices of inequality.
\textit{International Economic Review,} \textbf{29}, 147--156.

\hang
\textsc{Chakravarty, S. R.} (2007)
A deprivation-based axiomatic characterization of the absolute Bonferroni
index of inequality.
\textit{Journal of Economic Inequality}, \textbf{5}, 339--351.

\hang
\textsc{Chakravarty, S. R. \& Muliere, P.} (2004)
Welfare indicators: a review and new perspectives. 2. Measurement of poverty. \textit{Metron}, \textbf{62}, 247--281.

\hang
\textsc{Champernowne, D. G. \& Cowell,  F. A.} (1998)
\textit{Economic Inequality and Income Distribution.}
Cambridge: Cambridge University Press.

\hang
\textsc{Cobham, A. \& Sumner, A.} (2013a)
Putting the Gini back in the bottle? `The Palma' as a policy-relevant measure of inequality.
\textit{Working Paper 2013-5.} London: King's College.

\hang
\textsc{Cobham, A. \& Sumner, A.} (2013b)
Is it all about the tails? The Palma measure of income inequality.
\textit{\textit{Working Paper 343.}} Washington DC: Center for Global Development.

\hang
\textsc{Cobham, A. \& Sumner, A.} (2014)
In inequality all about the tails?: The Palma measure of income inequality. \textit{Significance}, \textbf{11}, 10--13.  

\hang
\textsc{Cowell, F. A.} (2011)
\textit{Measuring Inequality}, 3rd edn.
Oxford: Oxford University Press.

\hang
\textsc{Cruz, M.} (2009)
\textit{The Solvency II Handbook.}
London: Risk Books.

\hang
\textsc{Denneberg, D.} (1990)
Premium calculation: why standard deviation should be replaced by absolute deviation.
\textit{ASTIN Bulletin}, \textbf{20}, 181--190.

\hang
\textsc{De Vergottini, M.} (1940)
Sul significato di alcuni indici di concentrazione.
\textit{Giornale degli Economisti e Annali di Economia}, \textbf{11}, 317--347.

\hang
\textsc{Donaldson, D. \&   Weymark, J. A.} (1980)
A single-parameter generalization of the Gini indices of inequality.
\textit{Journal of Economic Theory}, \textbf{22}, 67--86.

\hang
\textsc{Donaldson, D. \&  Weymark, J. A.} (1983)
Ethically flexible Gini indices for income distributions in the continuum.
\textit{Journal of Economic Theory}, \textbf{29}, 353--358.

\hang
\textsc{Druckman, A. \& Jackson, T.} (2008)
Measuring resource inequalities: The concepts and methodology for an area-based Gini coefficient.
\textit{Ecological Economics}, \textbf{65}, 242--252.

\hang
\textsc{Duclos, J. Y.} (2000)
Gini indices and the redistribution of income.
\textit{International Tax and Public Finance}, \textbf{7}, 141--162.

\hang
\textsc{Fort, R. \& Quirk, J.} (1995)
Cross-subsidization, incentives, and outcomes in professional team sports leagues.
\textit{Journal of Economic Literature}, \textbf{33}, 1265--1299.

\hang
\textsc{Furman, E. \& Zitikis, R.} (2008)
Weighted premium calculation principles.
\textit{Insurance: Mathematics and Economics}, \textbf{42}, 459--465.

\hang
\textsc{Furman, E. \& Zitikis, R.} (2009)
Weighted pricing functionals with applications to insurance: an overview.
\textit{North American Actuarial Journal}, \textbf{13}, 483--496.

\hang
\textsc{Gastwirth, J. L.} (1971)
A general definition of the Lorenz curve.
\textit{Econometrica}, \textbf{39}, 1037--1039.

\hang
\textsc{Gastwirth, J. L.} (2014)
Median-based measures of inequality: reassessing the increase in income inequality in the U.S. and Sweden.
\textit{Journal of the IAOS}, \textbf{30}, 311--320.

\hang
\textsc{Gini, C.} (1912)
\textit{Variabilit{\`a} e Mutabilit{\`a}: Contributo allo Studio delle Distribuzioni e delle Relazioni Statistiche}. Bologna: Tipografia di Paolo Cuppini.

\hang
\textsc{Gini, C.} (1914)
On the measurement of concentration and variability of characters (English translation from Italian by Fulvio de Santis) \textit{Metron}, \textbf{63}, 3--38.

\hang
\textsc{Gini, C.} (1921)
Measurement of inequality of incomes. \textit{Economic Journal}, \textbf{31}, 124--126

\hang
\textsc{Giorgi, G. M.} (1990)
Bibliographic portrait of the Gini concentration ratio. \textit{Metron}, \textbf{48}, 183--221.

\hang
\textsc{Giorgi, G. M.} (1993)
A fresh look at the topical interest of the Gini  concentration ratio.
\textit{Metron}, \textbf{51}, 83--98.

\hang
\textsc{Giorgi, G. M.} (1998)
Concentration index, Bonferroni.
\textit{Encyclopedia of Statistical Sciences} (S. Kotz, D. L. Banks \& C. B. Read eds), vol.~2. New York: Wiley, pp. 141--146.

\hang
\textsc{Giorgi, G. M. \&   Crescenzi, M.} (2001)
A look at the Bonferroni inequality measure in a reliability framework.
\textit{Statistica}, \textbf{91}, 571--583.

\hang
\textsc{Giorgi, G. M. \&   Nadarajah, S.} (2010)
Bonferroni and Gini indices for various parametric families of distributions.
\textit{Metron}, \textbf{68}, 23--46.

\hang
\textsc{Giovagnoli, A. \& Wynn, H.} (2012)
$(U,V)$-ordering and a duality theorem for risk aversion and Lorenz-type orderings. LSE Philosophy Papers. London: London School of Economics and Political Science.

\hang
\textsc{Greselin, F.} (2014) 
More equal and poorer, or richer but more unequal? 
\textit{Economic Quality Control}, \textbf{29}, 99--117.

\hang
\textsc{Greselin, F., Pasquazzi, L. \& Zitikis, R.} (2013)
Contrasting the Gini and Zenga indices of economic inequality. 
\textit{Journal of Applied Statistics}, \textbf{40}, 282--297.

\hang
\textsc{Greselin, F., Puri, M. L. \& Zitikis, R.} (2009)
$L$-functions, processes, and statistics in measuring economic inequality and actuarial risks.
\textit{Statistics and Its Interface}, \textbf{2}, 227--245.

\hang
\textsc{Hand, D. J.} (2001)
Modelling consumer credit risk.
\textit{IMA Journal of Management Mathematics}, \textbf{12}, 139--155.

\hang
\textsc{Harsanyi, J. C.}  (1953)
Cardinal utility in welfare economics and in the theory of risk-taking.  \textit{Journal of Political Economy}, \textbf{61}, 434--435.

\hang
\textsc{Imedio-Olmedo, L. J., B\'arcena-Mart\'in, E. \& Parrado-Gallardo, E. M.} (2011) A class of Bonferroni inequality indices.
\textit{Journal of Public Economic Theory}, \textbf{13}, 97--124.

\hang
\textsc{Jaworski, P., Durante, F. \& H\"{a}rdle, W. K.} (eds) (2013) \textit{Copulae in Mathematical and Quantitative Finance.}
Berlin: Springer.

\hang
\textsc{Jaworski, P., Durante, F., H\"{a}rdle, W. \& Rychlik, T.} (eds) (2010)
\textit{Copula Theory and Its Applications.}
Berlin: Springer.

\hang
\textsc{Jones, B. L. \& Zitikis, R.} (2003)
Empirical estimation of risk measures and related quantities.
\textit{North American Actuarial Journal}, \textbf{7}, 44--54.

\hang
\textsc{Kakwani, N. C.} (1980a)
\textit{Income Inequality and Poverty: Methods of Estimation and Policy Applications}.
New York: Oxford University Press.

\hang
\textsc{Kakwani, N.} (1980b)
On a class of poverty measures.
\textit{Econometrica}, \textbf{48}, 437--446.

\hang
\textsc{Kakwani, N. C. \& Podder, N.} (1973)
On the estimation of Lorenz curves from grouped observations.
\textit{International Economic Review}, \textbf{14}, 278--292.

\hang
\textsc{Kenworthy, L. \& Pontusson, J.} (2005)
Rising inequality and the politics of redistribution
in affluent countries.
\textit{Perspectives on Politics}, \textbf{3}, 449--471.

\hang
\textsc{Kleiber, C. \&   Kotz, S.} (2003)
\textit{Statistical Size Distributions in Economics and Actuarial Sciences}. Hoboken: Wiley.

\hang
\textsc{Korpi, W. \& Palme, J.} (1998)
The paradox of redistribution and strategies of equality: welfare state institutions, inequality, and poverty in the Western countries.
\textit{American Sociological Review}, \textbf{63}, 661--687.

\hang
\textsc{Kovacevic, M.} (2010)
Measurement of inequality in human development -- a review.
\textit{Human Development Research Paper 2010/35}. New York:
United Nations Development Programme, United Nations.

\hang
\textsc{Krzanowski, W. J. \& Hand, D. J.} (2009)
\textit{ROC Curves for Continuous Data.}
London: CRC Press.

\hang
\textsc{Lambert, P. J.} (2001)
\textit{The Distribution and Redistribution of Income}, 3rd edn.
Manchester: Manchester University Press.

\hang
\textsc{Li, H. \& Li, X.} (eds) (2013)
\textit{Stochastic Orders in Reliability and Risk: In Honor of Professor Moshe Shaked}. New York: Springer.

\hang
\textsc{Lorenz, M. O.} (1905)
Methods of measuring the concentration of wealth.
\textit{Publications of the American Statistical Association}, \textbf{9}, 209--219.

\hang
\textsc{Machina, M.} (1987)
Choice under uncertainty: problems solved and unsolved.
\textit{Economic Perspectives}, \textbf{1}, 121--154.

\hang
\textsc{Machina, M.} (2008)
Non-expected utility theory.
\textit{The New Palgrave Dictionary of Economics}
(S. N. Durlauf \& L. E. Blume eds), 2nd edn.
New York: Palgrave Macmillan, pp. 74--84.

\hang
\textsc{Manasis, V.,  Avgerinou, V.,  Ntzoufras, I. \& Reade, J. J.} (2013) Quantification of competitive balance in European football: development of specially designed indices.
\textit{IMA Journal of Management mathematics}, \textbf{24}, 363--375.

\hang
\textsc{Manasis, V. \&  Ntzoufras, I.} (2014)
Between-seasons competitive balance in European football: review of existing and development of specially designed indices.
\textit{Journal of Quantitative Analysis in Sports}, \textbf{10}, 139--152.

\hang
\textsc{McNeil, A. J., Frey, R. \&   Embrechts, P.} (2005)
\textit{Quantitative Risk Management}. Princeton: Princeton University Press.

\hang
\textsc{Mimoto, N. \&  Zitikis, R.} (2008)
The Atkinson index, the Moran statistic, and testing exponentiality.
\textit{Journal of the Japan Statistical Society}, 38, 187--205.

\hang
\textsc{Meucci, A.} (2007)
\textit{Risk and Asset Allocation}, 3rd printing.
Berlin: Springer.

\hang
\textsc{Muliere, P. \&   Scarsini, M.} (1989)
A note on stochastic dominance and inequality measures.
\textit{Journal of Economic Theory}, \textbf{49}, 314--323.

\hang
\textsc{Nelsen, R. B.} (2006)
\textit{An Introduction to Copulas,} 2nd edn. New York: Springer.

\hang
\textsc{Nyg{\aa}rd, F. \& Sandstr\"{o}m, A.} (1981)
\textit{Measuring Income Inequality.}
Stockholm: Almqvist and Wiksell.

\hang
\textsc{Oladosu, G. \& Rose, A.} (2007)
Income distribution impacts of climate change mitigation policy in the Susquehanna River Basin Economy.
\textit{Energy Economics}, \textbf{29},  520--544.

\hang
\textsc{Ostry, J. D., Berg, A. \& Tsangarides, C. G.} (2014)
Redistribution, Inequality, and Growth.
\textit{IMF Staff Discussion Note SDN/14/02.}
Washington DC: Research Department, International Monetary Fund.

\hang
\textsc{Palma, J. G.} (2006)
Globalizing inequality: `centrifugal' and `centripetal' forces at work.
\textit{DESA Working Paper No 35.} New York: Department of Economics and Social Affairs, United Nations.

\hang
\textsc{Pf\"{a}hler, W.} (1990)
Redistributive effect of income taxation: decomposing tax base and tax rates effects.
\textit{Bulletin of Economic Research}, \textbf{42}, 121--129.

\hang
\textsc{Pflug, G. Ch. \& R\"{o}misch, W.} (2007)
\textit{Modeling, Measuring and Managing Risk}.
Singapore: World Scientific.

\hang
\textsc{Pietra, G.} (1915)
On the relationship between variability indices (Note I).
(English translation from Italian by P. Brutti \& S. Gubbiotti) \textit{Metron}, \textbf{72}, 5--16.

\hang
\textsc{Piketty, T.} (2014)
\textit{Capital in the Twenty-First Century.}
Cambridge: Harvard University Press.


\hang
\textsc{Puppe, C.} (1991)
\textit{Distorted Probabilities and Choice under Risk.}
Berlin: Springer.


\hang
\textsc{Rawls, J.} (1971)
\textit{A Theory of Justice.} Cambridge: Harvard University Press.

\hang
\textsc{Roemer, J. E.} (2013)
Economic development as opportunity equalization.
\textit{World Bank Economic Review}, \textbf{28}, 189--209.

\hang
\textsc{Quiggin, J.} (1982)
A theory of anticipated utility.
\textit{Journal of Economic Behavior and Organization}, \textbf{3}, 323--343.

\hang
\textsc{Quiggin, J.} (1993)
\textit{Generalized Expected Utility Theory: The Rank-Dependent Model.}
Kluwer: Dordrecht.

\hang
\textsc{Sadoulet, E. \& de Janvry, A.} (1995)
\textit{Quantitative Development Policy Analysis.}
Baltimore: John Hopkins University Press.

\hang
\textsc{Sandstr\"{o}m, A.} (2010)
\textit{Handbook of Solvency for Actuaries and Risk Managers: Theory and Practice.}
Boca Raton: Chapman and Hall.

\hang
\textsc{Sarabia, J. M.} (2008)
Parametric Lorenz curves: models and applications.
\textit{Modeling Income Distributions and Lorenz Curves}
(D. Chotikapanich ed.). Berlin: Springer, pp. 167--190.

\hang
\textsc{Sarabia, J. M., Prieto, F. \& and Sarabia, M.} (2010)
Revisiting a functional form for the Lorenz curve.
\textit{Economics Letters}, \textbf{107}, 249--252.

\hang
\textsc{Schmeidler, D.} (1986)
Integral representation without additivity.
\textit{Proceedings of the American Mathematical Society}, \textbf{97}, 255--261.

\hang
\textsc{Schmeidler, D.} (1989)
Subjective probability and expected utility without additivity.
\textit{Econometrica}, \textbf{57}, 571--587.

\hang
\textsc{Sen, A.} (1983)
Poor, relatively speaking.
\textit{Oxford Economic Papers}, \textbf{35}, 153--169.

\hang
\textsc{Sen, A.} (1997)
\textit{On Economic Inequality}, expanded edition with a substantial annexe by J. E. Foster and A. Sen.
Oxford: Clarendon Press.

\hang
\textsc{Sen, A.} (1998)
\textit{Choice, Welfare and Measurement}, 2nd printing.
Cambridge: Harvard University Press.

\hang
\textsc{Shaked, M. \& Shanthikumar, J. G.} (2007)
\textit{Stochastic Orders}. New York: Springer.

\hang
\textsc{Shorrocks, A.} (1978)
Income inequality and income mobility.
\textit{Journal of Economic Theory}, \textbf{19}, 376--393.

\hang
\textsc{Silber, J.} (ed.) (1999)
\textit{Handbook on Income Inequality Measurement.}
Boston: Kluwer.

\hang
\textsc{Slemrod, J.} (1992)
Taxation and inequality: a time-exposure perspective. \textit{Tax Policy and the Economy} (J. M. Poterba ed.), vol. 6. Chicago: University of Chicago Press, pp. 105--127.

\hang
\textsc{Sordo, M. A. \& Su\'{a}rez-Llorens, A.} (2011)
Stochastic comparisons of distorted variability measures.
\textit{Insurance: Mathematics and Economics}, \textbf{49}, 11--17.

\hang
\textsc{Sordo, M. A., Navarro, J. \& Sarabia, J. M.} (2014)
Distorted Lorenz curves: models and comparisons.
\textit{Social Choice and Welfare}, \textbf{42}, 761--780.

\hang
\textsc{Tarsitano, A.} (1990)
The Bonferroni index of income inequality.
\textit{Income and Wealth Distribution, Inequality and Poverty} (C. Dagum \& M. Zenga eds). New York: Springer, pp. 228--242.

\hang
\textsc{Tarsitano, A.} (2004)
A new class of inequality measures based on a ratio of $L$-statistics. \textit{Metron}, \textbf{62}, 137--160.

\hang
\textsc{Thompson, W. A. Jr.} (1976)
Fisherman's luck.
\textit{Biometrics}, \textbf{32}, 265--271.

\hang
\textsc{Van De Ven, J., Creedy, J. \& Lambert, P. J.} (2001)
Close equals and calculation of the vertical, horizontal and reranking effects of taxation.
\textit{Oxford Bulletin of Economics and Statistics}, \textbf{63}, 381--394.

\hang
\textsc{Vergnaud, J. C.} (1997)
Analysis of risk in a non expected utility framework and application to the optimality of the deductible.
\textit{Revue Finance}, \textbf{18}, 155--167.

\hang
\textsc{Wang, S.} (1995)
Insurance pricing and increased limits ratemaking by proportional hazards transforms.
\textit{Insurance: Mathematics and Economics}, \textbf{17}, 43--54.

\hang
\textsc{Wang, S.} (1998)
An actuarial index of the right-tail risk.
\textit{North American Actuarial Journal}, \textbf{2}, 88--101.

\hang
\textsc{Wang, S. S. \& Young, V. R.} (1998)
Ordering risks: expected utility theory
versus Yaari's dual theory of risk.
\textit{Insurance: Mathematics and Economics}, \textbf{22}, 145--161.

\hang
\textsc{Wang, S. S., Young, V. R. \& Panjer, H. H.}  (1997)
Axiomatic characterization of insurance prices.
\textit{Insurance: Mathematics and Economics}, \textbf{21}, 173-183

\hang
\textsc{Weymark, J. A.}  (1981)
Generalized Gini inequality indices.
\textit{Mathematical Social Sciences}, \textbf{1}, 409--430.

\hang
\textsc{Weymark, J.}  (2003)
Generalized Gini indices of equality of opportunity.
\textit{Journal of Economic Inequality}, \textbf{1}, 5-24.

\hang
\textsc{Wirch, J. L. \& Hardy, M. R.} (1999)
A synthesis of risk
measures for capital adequacy.
\textit{Insurance: Mathematics and Economics}, \textbf{25}, 337--347.

\hang
\textsc{Yaari, M. E.} (1987)
The dual theory of choice under risk.
\textit{Econometrica}, \textbf{55}, 95--115.

\hang
\textsc{Yitzhaki, S.} (1979)
Relative deprivation and the Gini coefficient.
\textit{Quarterly Journal of Economics}, \textbf{93}, 321--324.

\hang
\textsc{Yitzhaki, S.} (1982)
Stochastic dominance, mean variance, and Gini's mean difference.
\textit{American Economic Review}, \textbf{72}, 178--185.

\hang
\textsc{Yitzhaki, S.} (1983)
On an extension of the Gini inequality index.
\textit{International Economic Review}, \textbf{24}, 617--628.

\hang
\textsc{Yitzhaki, S.} (1994)
On the progressivity of commodity taxation.
\textit{Models and Measurement of Welfare and Inequality} 
(W. Eichhorn ed.).
Berlin: Springer, pp. 448--466.

\hang
\textsc{Yitzhaki, S.} (1998) More than a dozen alternative ways of spelling Gini. \textit{Research on Economic Inequality}, 8, 13--30.

\hang
\textsc{Yitzhaki, S.} (2003)
Gini's mean difference: A superior measure of variability for non-normal distributions.
\textit{Metron}, \textbf{51}, 285--316.

\hang
\textsc{Yitzhaki, S. \& Schechtman, E.} (2013)
\textit{The Gini Methodology: A Primer on a Statistical Methodology.}
New York: Springer.

\hang
\textsc{Zenga, M.} (2007)
Inequality curve and inequality index based on the ratios between lower and upper arithmetic means.
\textit{Statistica \& Applicazioni}, \textbf{5}, 3--27.

\hang
\textsc{Zenga, M.} (2010)
Mixture of Polisicchio's truncated Pareto distributions with beta weights.
\textit{Statistica \& Applicazioni}, \textbf{8}, 3--25.

\hang
\textsc{Zenga M., Pasquazzi, L., Polisicchio, M. \& Zenga, M.} (2011)
More on M.~M.~Zenga's new three-parameter distribution for non-negative variables.
\textit{Statistica \& Applicazioni},  \textbf{9}, 5--33.

\hang
\textsc{Zenga, M., Pasquazzi, L. \& Zenga, M.} (2012)
First applications of a new three-parameter distribution for non-negative variables.
\textit{Statistica \& Applicazioni}, \textbf{10}, 131--147.

\hang
\textsc{Zitikis, R.} (2002)
Analysis of indices of economic inequality from a mathematical point of view. (Invited Plenary Lecture at the 11th Indonesian Mathematics Conference, State University of Malang, Indonesia.) \textit{Matematika}, \textbf{8}, 772--782.

\hang
\textsc{Zitikis, R.} (2003)
Asymptotic estimation of the $E$-Gini index.
\textit{Econometric Theory}, \textbf{19}, 587--601.

\appendix

\section{\normalsize Appendix: Technicalities}
\label{appendix}

\paragraph{\sc Proof of equations (\ref{eq-dw-0b}).}

Since the relative-value function is $v(x,y)=1-x/y$, we have
\begin{equation}\label{eq-dw-1}
\mathrm{DWK}_{F,h}
= 1- {1\over \mu_F }\int_0^1 \mathrm{LCE}_F(p) f(p) \, dp,
\end{equation}
where $f(p)$ is the density function of the gamble $\pi_h $ defined by equation (\ref{choice-dw}). The following are straightforward calculations:
\begin{align*}
\int_0^1 \mathrm{LCE}_F(p)f(p) \, dp&=   \int_0^1 \frac{1}{p} \bigg ( \int_0^p F^{-1}(t) \, dt \bigg )  f(p) \,dp \\
&= \int_0^1  \frac{1}{p} \bigg (\int_0^1 \mathbf{1}\{t \leq p \} F^{-1}(t) \, dt \bigg )  f(p) \,dp \\
&= \int_0^1  F^{-1}(t) \left( \int_0^1 \frac{1}{p}  \, \mathbf{1} \{t \leq p \} \, f(p) \,dp \right) \, dt\\
&= \int_0^1  F^{-1}(t) \left( \int_t^1 \frac{1}{p}  \, f(p) \,dp \right) \, dt \\
&={1\over 1- h'(0)} \int_0^1  F^{-1}(t) \big ( h'(1-t) -h'(0) \big ) \, dt \\
&={1\over 1- h'(0)} \bigg ( \int_0^1  F^{-1}(t) h'(1-t) \, dt  -h'(0) \mu_F \bigg ).
 \end{align*}
Combining this result with equation (\ref{eq-dw-1}), we obtain the first equation of (\ref{eq-dw-0b}).  Since
\begin{align}
\int_0^1  F^{-1}(t) h'(1-t) \, dt 
&=\int_0^\infty \bigg (\int_0^1 \mathbf{1}\{ F^{-1}(t)>x\} h'(1-t) \, dt \bigg ) dx
\notag
\\
&=\int_0^\infty \bigg (\int_0^1 \mathbf{1}\{ t>F(x)\} h'(1-t) \, dt \bigg ) dx
\notag
\\
&=\int_0^\infty \bigg (\int_{F(x)}^1 h'(1-t) \, dt \bigg ) dx
\notag
\\
&=\int_0^\infty h(1-F(x)) \, dx,
\label{eq-integral}
\end{align}
we have the second equation of  (\ref{eq-dw-0b}).

\paragraph{\sc Proof of equations (\ref{eq-w-0b}).}

Since the relative-value function is $v(x,y)=x/y-1$, we have
\begin{equation}\label{eq-w-1}
\mathbf{E}[v(\mathrm{UCE}_F(\pi_g),\mu_F)]= {1\over \mu_F }\int_0^1 \mathrm{UCE}_F(p) f(p) \, dp -1 ,
\end{equation}
where $f(p)$ is the density function of the gamble $\pi_g $ defined by equation (\ref{choice-wang}). The following are straightforward calculations:
\begin{align*}
\int_0^1 \mathrm{UCE}_F(p)f(p) \, dp&=   \int_0^1 \frac{1}{1-p} \bigg ( \int_p^1 F^{-1}(t) \, dt \bigg )  f(p) \,dp \\
&= \int_0^1\bigg ( \int_0^1 \mathbf{1}\{t \geq p \} F^{-1}(t) \, dt \bigg ) \frac{f(p)} {1-p}  \,dp \\
&= \int_0^1  F^{-1}(t) \left( \int_0^1  \mathbf{1} \{t \geq p \} \,\frac{f(p)} {1-p} \,dp \right) \, dt\\
&= \int_0^1  F^{-1}(t) \left( \int_0^t \frac{f(p)}{1-p} \,dp \right) \, dt .
\end{align*}
Applying definition (\ref{choice-wang}) of the density function $f(p)$, we obtain
\begin{align}
\int_0^1 \mathrm{UCE}(p)f(p) \, dp
&={1\over 1- g'(1)} \int_0^1  F^{-1}(t) \big ( g'(1-t)-g'(1) \big ) \, dt
\notag
\\
&={1\over 1- g'(1)}\bigg ( \int_0^1  F^{-1}(t) g'(1-t)\, dt - g'(1)\mu_F \bigg ).
\label{new-1}
 \end{align}
Combining equations (\ref{new-1}) and (\ref{eq-w-1}), we obtain the first equation of (\ref{eq-w-0b}). Using equation (\ref{eq-integral}) with $g$ instead of $h$, we arrive at the second equation of  (\ref{eq-w-0b}).

\begin{note}\rm
From the mathematical point of view, equation (\ref{eq-integral}) is elementary, but it was a pivotal observation that allowed Jones and Zitikis~(2003) to initiate the development of statistical inference for the Wang (or distortion) risk measure. Since then, numerous statistical results have appeared on risk measures: parametric and non-parametric, light- and heavy-tailed cases have been explored in great detail by many authors.
\end{note}

\paragraph{\sc Proof of Property \ref{prop-0}.}

Part (\ref{prop-i}) follows from the fact that if $X=d$ for any  constant $d>0$, then $F_{X}^{-1}(p)=d$ and so $\mathrm{UCE}_X(p)= \mathrm{LCE}_X(p)$ for every $p\in (0,1)$. Part (\ref{prop-ii}) follows from the fact that if $d>0$, then $F_{dX}^{-1}(p)=dF_{X}^{-1}(p)$ and so $\mathrm{UCE}_{dX}(p)/ \mathrm{LCE}_{dX}(p)=\mathrm{UCE}_{X}(p)/ \mathrm{LCE}_{X}(p)$ for every $p\in (0,1)$. Part (\ref{prop-iii}) follows from the fact that $F_{X+d}^{-1}(p)=F_{X}^{-1}(p)+d$ for every $d$, and so the bound $\mathrm{LCE}_{X}(p)\le \mathrm{UCE}_{X}(p)$ together with the assumed positivity of $d$ imply 
\[
{\mathrm{UCE}_{X+d}(p)\over \mathrm{LCE}_{X+d}(p)}
={\mathrm{UCE}_{X}(p)+d\over \mathrm{LCE}_{X}(p)+d}
\le {\mathrm{UCE}_{X}(p)\over \mathrm{LCE}_{X}(p)} .
\]
The latter bound is equivalent to $\mathrm{R}_{X+d}(p) \le \mathrm{R}_X(p)$ for every $p\in (0,1)$, which establishes the bound $\mathrm{R}_{X+d} \le \mathrm{R}_X$.

\end{document}